\documentclass[12pt,a4p,epsfig,float,amssymb]{article}
\usepackage{epsfig}
\usepackage{cite}

\def\K+K- {$\mbox{K}^{+}\mbox{K}^{-}$}

\def\y32 {$y_{32}$}

\def\epm {\mbox{$\mathrm{e}^+\mathrm{e}^-$}}
\def\bec {BEC}
\def\bose {Bose-Einstein correlations}

\newcommand{\epem}{\ensuremath{\mathrm{e}^+\mathrm{e}^-}}

\newcommand{\Zz}{\ensuremath{{\mathrm{Z}^0}}}

\newcommand{\WW}{\ensuremath{\mathrm{W}^+\mathrm{W}^-}}

\newcommand{\eeWW}{\ensuremath{\epem\rightarrow\WW}}
\newcommand{\qq}{\ensuremath{\mathrm{q\overline{q}}}}

\newcommand{\lnu}{\ensuremath{\ell\overline{\nu}_{\ell}}}

\newcommand{\enu}{\ensuremath{\mathrm{e\overline{\nu}_{e}}}}
\newcommand{\mnu}{\ensuremath{\mu\overline{\nu}_{\mu}}}

\newcommand{\tnu}{\ensuremath{\tau\overline{\nu}_{\tau}}}

\newcommand{\WWqqqq}{\ensuremath{\WW\rightarrow\qq\qq}}
\newcommand{\WWqqln}{\ensuremath{\WW\rightarrow\qq\lnu}}
\newcommand{\WWqqen}{\ensuremath{\WW\rightarrow\qq\enu}}
\newcommand{\WWqqmn}{\ensuremath{\WW\rightarrow\qq\mnu}}
\newcommand{\WWqqtn}{\ensuremath{\WW\rightarrow\qq\tnu}}

\newcommand{\Wenu}{\ensuremath{\epem \rightarrow \mathrm{W}\enu}}
\newcommand{\singW}{\ensuremath{\mathrm{W}\enu}}
\newcommand{\ZZqqqq}{\ensuremath{\Zz \Zz \rightarrow\qq\qq}}
\newcommand{\ZZqqll}{\ensuremath{\Zz \Zz \rightarrow\qq\ell \overline {\ell}}}

\newcommand{\Zqq}{\ensuremath{(\Zz/\gamma)^{*}\rightarrow\qq}}

\newcommand{\fpi}{\ensuremath{f_{\pi}(Q)}}
\newcommand{\csemi}{\ensuremath{{C^{\mathrm{semi}}(Q)}}}
\newcommand{\cwwqqlv}{\ensuremath{{C^{{\mathrm {\qq}}}(Q)}}}
\newcommand{\csame}{\ensuremath{{C^{\mathrm {same}}(Q)}}}
\newcommand{\czstar}{\ensuremath{{C^{{\mathrm {Z}}^{*}}(Q)}}}
\newcommand{\cwwqqqq}{\ensuremath{{C^{{\mathrm {\qq \qq}}}(Q)}}}
\newcommand{\cstarone}{\ensuremath{{C^{{\mathrm {Z}}^{*}}_{\mathrm{had}}(Q)}}}
\newcommand{\cdiff}{\ensuremath{{C^{\mathrm {diff}}(Q)}}}
\newcommand{\cqcd}{\ensuremath{{C^{\mathrm {non-rad.}}(Q)}}}
\newcommand{\chad}{\ensuremath{{C^{\mathrm {had}}(Q)}}}
\newcommand{\psamesemi}{\ensuremath{{P^{\mathrm
        {W}}_{\mathrm{semi}}(Q)}}}
\newcommand{\psamehad}{\ensuremath{{P^{\mathrm
        {same}}_{\mathrm{had}}(Q)}}}
\newcommand{\pzstarhad}{\ensuremath{{P^{\mathrm
        {Z^{*}}}_{\mathrm{had}}(Q)}}}

\newcommand{\psameqcd}{\ensuremath{{P^{\mathrm {same}}_{\mathrm{non-rad}}(Q)}}}
\newcommand{\pzstarqcd}{\ensuremath{{P^{\mathrm{Z^{*}}}_{\mathrm{non-rad}}(Q)}}}
\newcommand{\pwwqqqq}{\ensuremath{{P^{\mathrm {WW}}_{\mathrm{had}}(Q)}}}
\newcommand{\pwwqqln}{\ensuremath{{P^{\mathrm {W}}_{\mathrm{semi}}(Q)}}}
\newcommand{\pwwnorad}{\ensuremath{{P^{\mathrm {WW}}_{\mathrm{non-rad}}(Q)}}}
\newcommand{\pznorad}{\ensuremath{{P^{\mathrm {Z^{*}}}_{\mathrm {non-rad}}(Q)}}}

\newcommand{\lamsame}{\ensuremath{{\lambda^{\mathrm{ same}}}}}
\newcommand{\lamdiff}{\ensuremath{{\lambda^{\mathrm {diff}}}}}
\newcommand{\lamzstar}{\ensuremath{{\lambda^{\mathrm Z^{*}}}}}
\newcommand{\lamwwqqqq}{\ensuremath{{\lambda^{\mathrm \WWqqqq}}}}
\newcommand{\lamwwqqln}{\ensuremath{{\lambda^{\mathrm \WWqqln}}}}

\newcommand{\deldiff}{\ensuremath{{\delta^{\mathrm {diff}}}}}

\newcommand{\epsdiff}{\ensuremath{{\epsilon^{\mathrm {diff}}}}}

\newcommand{\lll}{\bigg/}

\newcommand{\Jetset}{\mbox{J{\sc etset}}}
\newcommand{\Koralw}{\mbox{K{\sc oralw}}}

\newcommand{\Pythia}{\mbox{P{\sc ythia}}}

\newcommand{\Herwig}{\mbox{H{\sc erwig}}}

\newcommand{\dedx}{\ensuremath{\mathrm{d}E/\mathrm{d}x}}

\newcommand{\rootsprime}{\ensuremath{\sqrt{s^\prime}}}

\newcommand{\Zgamma}{\ensuremath{(\Zz/\gamma)^{*}\,\,}}

\newcommand {\ee}         {\ensuremath{\mathrm{e}^+ \mathrm{e}^-}}

\newcommand {\tautau}     {\ensuremath{\tau^+ \tau^-}}

\begin{document}
\begin{titlepage}
\def\toprule{\noalign{\hrule \medskip}}
\def\midrule{\noalign{\medskip\hrule }}
\def\botrule{\noalign{\medskip\hrule }}
\flushbottom
\begin{center}
{\large\bf EUROPEAN LABORATORY FOR PARTICLE PHYSICS}
\end{center}
\bigskip
\begin{flushright}
CERN-EP/98-174 \\ 
28th October 1998  \\
\end{flushright}
\bigskip\bigskip\bigskip\bigskip\bigskip
\begin{center}
{\LARGE\bf Bose-Einstein Correlations in }
{\LARGE\bf \\ \boldmath{ \eeWW} at 172 and 183 GeV}  \\
\vspace{2mm}
{\LARGE\bf  }
\end{center}
\bigskip\bigskip
\begin{center}{\LARGE The OPAL Collaboration
}\end{center}\bigskip\bigskip
\bigskip\begin{center}

\begin{abstract}
\bose\ between like-charge pions are studied in 
hadronic final states produced by
\epem\ anni\-hi\-la\-tions at center-of-mass energies of 172 and 183 GeV. 
Three event samples are studied, 
each dominated by one of the processes \WWqqln , \WWqqqq , or  \Zqq.
After demonstrating the existence of \bose\ in W decays, an attempt is
made to determine \bose\ for pions originating from the same W boson
and from different W bosons, as well as for pions from \Zqq\ events. 
The following results are obtained for the individual chaoticity parameters $\lambda$,
assuming a common source radius R:
\begin{eqnarray}
\lamsame{} & = & 0.63 \pm  0.19 \pm  0.14 \nonumber , \\
\lamdiff{} & = & 0.22  \pm  0.53 \pm 0.14  \nonumber , \\
\lamzstar{} & = & 0.47  \pm  0.11 \pm 0.08  \nonumber ,\\
R\, & = & 0.92\pm0.09\pm 0.09 \ \mathrm{fm}  \nonumber. 
\end{eqnarray}
In each case, the first error is statistical and the second is systematic. At 
the current level of statistical precision it is not established whether
\bose\, between pions from different W bosons exist or not. 
\end{abstract}
\end{center}
\bigskip\bigskip\bigskip\bigskip
\bigskip\bigskip
\begin{center}{\large
Submitted to Euro. Phys. Jour. C 
}\end{center}

\newpage
\end{titlepage}

\begin{center}{\Large        The OPAL Collaboration
}\end{center}\bigskip

\begin{center}{
G.\thinspace Abbiendi$^{  2}$,
K.\thinspace Ackerstaff$^{  8}$,
G.\thinspace Alexander$^{ 23}$,
J.\thinspace Allison$^{ 16}$,
N.\thinspace Altekamp$^{  5}$,
K.J.\thinspace Anderson$^{  9}$,
S.\thinspace Anderson$^{ 12}$,
S.\thinspace Arcelli$^{ 17}$,
S.\thinspace Asai$^{ 24}$,
S.F.\thinspace Ashby$^{  1}$
% ...
}\end{center}\bigskip
\bigskip
$^{  1}$School of Physics and Astronomy, University of Birmingham,
Birmingham B15 2TT, UK
\newline
$^{  2}$Dipartimento di Fisica dell' Universit\`a di Bologna and INFN,
I-40126 Bologna, Italy
\newline
$^{  3}$Physikalisches Institut, Universit\"at Bonn,
D-53115 Bonn, Germany
\newline
$^{  4}$Department of Physics, University of California,
Riverside CA 92521, USA
\newline
$^{  5}$Cavendish Laboratory, Cambridge CB3 0HE, UK
\newline
$^{  6}$Ottawa-Carleton Institute for Physics,
Department of Physics, Carleton University,
Ottawa, Ontario K1S 5B6, Canada
\newline
$^{  7}$Centre for Research in Particle Physics,
Carleton University, Ottawa, Ontario K1S 5B6, Canada
\newline
$^{  8}$CERN, European Organisation for Particle Physics,
CH-1211 Geneva 23, Switzerland
\newline
$^{  9}$Enrico Fermi Institute and Department of Physics,
University of Chicago, Chicago IL 60637, USA
\newline
$^{ 10}$Fakult\"at f\"ur Physik, Albert Ludwigs Universit\"at,
D-79104 Freiburg, Germany
\newline
$^{ 11}$Physikalisches Institut, Universit\"at
Heidelberg, D-69120 Heidelberg, Germany
\newline
$^{ 12}$Indiana University, Department of Physics,
Swain Hall West 117, Bloomington IN 47405, USA
\newline
$^{ 13}$Queen Mary and Westfield College, University of London,
London E1 4NS, UK
\newline
$^{ 14}$Technische Hochschule Aachen, III Physikalisches Institut,
Sommerfeldstrasse 26-28, D-52056 Aachen, Germany
\newline
$^{ 15}$University College London, London WC1E 6BT, UK
\newline
$^{ 16}$Department of Physics, Schuster Laboratory, The University,
Manchester M13 9PL, UK
\newline
$^{ 17}$Department of Physics, University of Maryland,
College Park, MD 20742, USA
\newline
$^{ 18}$Laboratoire de Physique Nucl\'eaire, Universit\'e de Montr\'eal,
Montr\'eal, Quebec H3C 3J7, Canada
\newline
$^{ 19}$University of Oregon, Department of Physics, Eugene
OR 97403, USA
\newline
$^{ 20}$CLRC Rutherford Appleton Laboratory, Chilton,
Didcot, Oxfordshire OX11 0QX, UK
\newline
$^{ 22}$Department of Physics, Technion-Israel Institute of
Technology, Haifa 32000, Israel
\newline
$^{ 23}$Department of Physics and Astronomy, Tel Aviv University,
Tel Aviv 69978, Israel
\newline
$^{ 24}$International Centre for Elementary Particle Physics and
Department of Physics, University of Tokyo, Tokyo 113-0033, and
Kobe University, Kobe 657-8501, Japan
\newline
$^{ 25}$Institute of Physical and Environmental Sciences,
Brunel University, Uxbridge, Middlesex UB8 3PH, UK
\newline
$^{ 26}$Particle Physics Department, Weizmann Institute of Science,
Rehovot 76100, Israel
\newline
$^{ 27}$Universit\"at Hamburg/DESY, II Institut f\"ur Experimental
Physik, Notkestrasse 85, D-22607 Hamburg, Germany
\newline
$^{ 28}$University of Victoria, Department of Physics, P O Box 3055,
Victoria BC V8W 3P6, Canada
\newline
$^{ 29}$University of British Columbia, Department of Physics,
Vancouver BC V6T 1Z1, Canada
\newline
$^{ 30}$University of Alberta,  Department of Physics,
Edmonton AB T6G 2J1, Canada
\newline
$^{ 31}$Duke University, Dept of Physics,
Durham, NC 27708-0305, USA
\newline
$^{ 32}$Research Institute for Particle and Nuclear Physics,
H-1525 Budapest, P O  Box 49, Hungary
\newline
$^{ 33}$Institute of Nuclear Research,
H-4001 Debrecen, P O  Box 51, Hungary
\newline
$^{ 34}$Ludwigs-Maximilians-Universit\"at M\"unchen,
Sektion Physik, Am Coulombwall 1, D-85748 Garching, Germany
\newline
\bigskip\newline
$^{  a}$ and at TRIUMF, Vancouver, Canada V6T 2A3
\newline
$^{  b}$ and Royal Society University Research Fellow
\newline
$^{  c}$ and Institute of Nuclear Research, Debrecen, Hungary
\newline
$^{  d}$ and Department of Experimental Physics, Lajos Kossuth
University, Debrecen, Hungary
\newline
$^{  e}$ on leave of absence from the University of Freiburg
\newline

\section{Introduction}
\label{intro}
In reactions leading to hadronic final states
 \bose\ (\bec ) between identical bosons are well known.
These correlations lead to an enhancement of
the number of identical bosons over that of
non-identical bosons when
the two particles are close to each other in phase space.
Experimentally this effect 
was first observed for pions by Goldhaber et al.~\cite{goldhaber}.
For recent reviews see, 
for example,  reference~\cite{marcellini}.
In \epm\ annihilations at center-of-mass energies of 91 GeV, \bec\, have been observed for charged
pion pairs ~\cite{opalbe,becmult, alephbe, delphibe}, for $\mathrm{K}^0_{\mathrm{S}}
\mathrm{K}^0_{\mathrm{S}}$ pairs ~\cite{opalk0k0,opalk0k02,delphik0k0,alephk0k0} 
and also for $\mathrm{K}^{\pm}\mathrm{K}^{\pm}$ ~\cite{delphikplus}.

\par
In the present paper we report on an investigation of \bec\, for charged pions between \epm\
reactions at center-of-mass energies of 172 and 183 GeV, above the
threshold for W-pair production. 
The analysis is motivated by the question of whether \bec\, for pions from different W bosons exist or not. 
Theoretically this question is still not settled ~\cite{lund,nobec}. 
However, if such correlations do exist, this could bias significantly  the
measurement of the W boson mass in fully hadronic W-pair events~\cite{lund,vato,
jadach,bib-lonnblad}. The DELPHI collaboration has published a measurement, at $\sqrt{s}=172$ GeV, of \bec\ between
pions originating from two different W bosons~\cite{bib-delphidiff}, 
in which basically \bec\ in \WWqqln\ events were subtracted 
from those of \WWqqqq\ events.
The aim of the present analysis is to analyse \bec\ for
fully hadronic W-pair events (\WWqqqq ), semileptonic W-pair events (\WWqqln ), as well as
non-radiative \Zqq\ events. After having established \bec\ in hadronic W decays,
\bec\, are investigated separately for three classes of pions: 
those originating from the same W boson, those from different W bosons 
and those from non-radiative \Zqq\ events. 
Note that in this analysis, tracks are not assigned to jets or W-bosons and no kinematic
fits are needed. \par 

\bec\ between
identical bosons can be formally expressed in terms of the normalised
function
\begin{equation}
\label{eq_1}
C(Q) \ = \ \frac{\rho_2(p_1,p_2)}{\rho_1(p_1)\rho_1(p_2)} \ = \ \sigma
\frac{d^2\sigma}{dp_1dp_2}\lll\left\{\frac{d\sigma}{dp_1}
\frac{d\sigma}{dp_2}\right\} \ ,
\end{equation}
where $\sigma$ is the total boson production cross section,
$\rho_1(p_i)$ and $d\sigma/dp_i$ are the single-boson density in
momentum space and the inclusive cross section, respectively.
Similarly $\rho_2(p_1,p_2)$ and $d^2\sigma/dp_1dp_2$ are respectively
the density of the two-boson system and its inclusive cross section.
The product of the independent one-particle densities
$\rho_1(p_1)\rho_1(p_2)$ is referred to as the reference density
distribution, to which the measured two-particle distribution
is compared. The inclusive two-boson density $\rho_2(p_1,p_2)$ can be
written as:
\begin{equation}
\label{eq_rho2}
\rho_2(p_1,p_2) \ = \ \rho_1(p_1)\rho_1(p_2) + K_2(p_1,p_2) \ ,
\end{equation}
where $K_2(p_1,p_2)$ represents the two-body correlations. In the simple
case of two identical bosons the normalised density function $C(Q)$,
defined in Eq.~\ref{eq_1}, describes the two-body
correlations.
Thus one has
\begin{equation}
\label{eq_r2}
C(Q) \ = \ 1 + \stackrel{\sim}{K}_2(p_1,p_2) \ ,
\end{equation}
where $\stackrel{\sim}{K}_2(p_1,p_2) =
K_2(p_1,p_2)/[\rho_1(p_1)\rho_1(p_2)]$ is the normalised two-body
correlation term. Since \bec\ are present when the
bosons are close to one another in phase space, a natural choice is
to study them as a function of the Lorentz invariant variable $Q$ defined by
\[ Q^2 = -(p_1 - p_2)^2 = M^2_2 - 4\mu^2 \ ,\] 
which approaches zero as the identical bosons move closer in phase
space. Here $p_i$ is the four-momentum vector of the $i$th particle,
$\mu$ is the boson mass (here $m_{\pi}$) and $M^2_2$ is the invariant mass squared of
the two-boson system.
Ideally the reference sample should contain all correlations  
present in the sample 
used to measure $\rho(p_1,p_2)$, other
than the \bec\ , such as those due to energy, 
momentum and charge conservation, resonance decays and 
global event properties.
In this analysis, the reference is chosen to be  
a sample of unlike-charge pairs of pions from the same event. 
Since the presence of the resonances  $\omega$, $\mathrm{K^0_S}$, $\mathrm{\eta}$,
$\mathrm{\eta^{ \prime}}$, $\mathrm{\rho^0}$, $\mathrm{f_{0}}$ and $\mathrm{f_{2}}$ 
in the unlike-charge reference sample leads to kinematic
correlations which are not present in the like-charge sample,
the unlike-charge sample has to  be corrected for this
effect using simulated events. \par
Assuming a spherically symmetric pion source with a Gaussian radial distribution, the 
correlation function $C(Q)$ can be parametrised~\cite{goldhaber} by 
\begin{equation}
C(Q) = N  \, (1 + f_{\pi}(Q)\,\lambda\,
{\mathrm{e}} ^ {-Q^2 R^2})\,(1 + \delta 
\, Q +\epsilon \, Q^2 ), 
\label{eq-usedfun}
\end{equation}
where $R$  is the  radius of the source and 
$\lambda$ 
represents the strength of the correlation, 
with $0 \leq \lambda \leq 1$. 
A value of $\lambda=1$ corresponds 
to a fully chaotic source, while $\lambda =0$ 
corresponds to a completely coherent source without any \bec .
The function $f_{\pi}(Q)$ is the probability that a selected track pair is
really a pair of pions, as a function of $Q$. 
The additional empirical term 
\mbox{$(1 + \delta\, Q 
+ \epsilon \, Q^2 )$}
takes into account the behaviour of the correlation function
at high $Q$ values due to long-range particle correlations
(e.g. charge and energy conservation, phase-space constraints), and
$N$ is a normalisation factor. \par
The structure of the paper is as follows.
Section 2 contains a brief overview of the OPAL detector, the event and
track selections as well as Monte Carlo models.
In section 3 the analysis of the data is described. \bec\ 
are investigated for \Zqq{}, \WWqqln{} and \WWqqqq{} events. After
establishing \bec\ in hadronic W-events the chaoticity parameter 
for \bec\ between the decay products from the same  W, \lamsame{}, and
from different W bosons, \lamdiff{}, are determined.
Finally, section 4 summarises the results obtained.
 
\section{Experimental Details}

\subsection{The OPAL detector}
\label{sec-det}
A detailed description of the OPAL detector has been presented 
elsewhere~\cite{opaldet} and therefore only the features
relevant to this analysis are summarised here.  Charged particle
trajectories are reconstructed using the cylindrical central tracking
detectors which consist of a silicon microvertex detector, a high-precision 
gas vertex detector, a large-volume gas jet chamber and thin
$z$-chambers \footnote{The OPAL right-handed coordinate system is defined such
that the origin is at the geometric centre of the jet chamber, $z$ is
parallel to, and has positive sense along, the e$^-$ beam direction, $r$
is the coordinate normal to $z$, $\theta$ is the polar angle with respect
to +$z$ and $\phi$ is the azimuthal angle around $z$.}.
The entire central detector is contained within a solenoid
that provides an axial magnetic field of 0.435~T\@.  
The silicon microvertex detector consists of two layers of
silicon strip detectors, allowing to measure at least one hit per charged track in the
angular  region
$|\cos\theta|<0.93$.  It is surrounded by the vertex drift 
chamber, followed by the jet chamber, about 400~cm in
length and 185~cm in radius, that provides up to 159 space points per
track and also measures the ionisation energy loss of charged particles,
\dedx. With at least 130 charge samples along a track, a resolution
of $3.8 \%$ is achieved for the \dedx\ for minimum ionising pions 
in jets ~\cite{jetchamber, bib-dEdx}.  
The $z$-chambers, which considerably improve the measurement of
charged tracks in $\theta$, follow the jet chamber at large radius.
The combination of these chambers leads to a momentum
resolution of $\sigma_{p}/p^{2}=1.25 \times 10^{-3}$ (GeV/$c$)$^{-1}$.  
Track finding is nearly 100\% 
efficient within the angular region $|\cos \theta |<0.92$ . 
The mass resolution for $\mathrm{K^{0}_{S}} \rightarrow \pi^{+} \pi^{-}$,
related to the resolution in the correlation variable $Q$,
is found to be $\sigma = 7.0\pm 0.1$ MeV/$c^{2}$ ~\cite{opalk0k0}.
 
\subsection{Data selection} 
\label{sec-selection}

This study is carried out using data at \epm\ center-of-mass
energies of 172 GeV and 183 GeV with integrated luminosities of approximately  
10~pb$^{-1}$ and 57~pb$^{-1}$, respectively.  
Three mutually exclusive event samples are selected: a) 
{\em the  fully hadronic  event sample}, \WWqqqq, where both W bosons decay
hadronically;
b){\em the semileptonic event sample}, \WWqqln,
where one W decays hadronically and the other decays semileptonically ;
and c)
hadronic non-W events \Zqq, referred to here as the  
{\em the non-radiative \Zgamma event sample} in this analysis. 
Throughout this paper, a reference to W$^{+}$ or its decay products 
implicitly includes the charge conjugate states. 

\subsubsection{{\bf Selection of the fully hadronic event sample \boldmath{\WWqqqq}}}
The selection of fully hadronic \WWqqqq\ events is performed in two 
stages using a preselection based on cuts followed by a likelihood--based 
selection procedure. Fully hadronic decays, \WWqqqq\ are 
characterised by four or more energetic hadronic jets and little missing energy.
A preselection using kinematic variables
removes background predominantly from radiative \Zqq\ events. 
Events satisfying the preselection criteria are subjected to a likelihood
selection, which discriminates between signal and the remaining
four-jet-like QCD background. \par
At 172 GeV, several variables based on the characteristic four-jet-like
nature, momentum balance and jet angular structure, are used to distinguish \WWqqqq\  
events from the remaining background and to construct the
likelihood. 
The details of the selection at 172 GeV are described in appendix B of~\cite{wmass172}. 
The signal and background situation at 183 GeV is similar to the one
at 172 GeV. For this reason, no new selection strategy was developed and the event
selection at 183 GeV is just a reoptimised version of the selection at 172 GeV.
The details of the selection at 183 GeV are described in~\cite{ww183}. At
183 GeV, no cut was applied against \Zz \Zz events. \par
Overall, there is a background of $11.6\%$ from \Zqq\ events  
and a contribution of $2.1\%$ from 
$\ee \rightarrow \ZZqqqq $ events. 
No selection for \WWqqqq\ events is applied to events 
selected as \WWqqln\ events. \par

\subsubsection{{\bf Selection of the semileptonic event sample \boldmath{\WWqqln}}}
\WWqqen\ and \WWqqmn\ events are characterised by two 
well-separated hadronic jets, a high-momentum lepton and 
missing momentum due to the unobserved neutrino.
In \WWqqtn\ the $\tau$ lepton gives rise to a low-multiplicity jet
consisting of one or three tracks.
The tracks from $\tau$ decay are not used in in the \bec\ studies.
Cuts are applied to reduce the background from radiative \Zqq\ events. 
A likelihood is formed using kinematic variables and 
characteristics of the lepton candidate to further suppress the background from \Zqq\ events. 
The details of the selection at 172 GeV are given in appendix A of~\cite{wmass172}. \par
The \WWqqln\ event selection for the 183~GeV data is a modified 
version of the 172~GeV selection.
At 183~GeV, a looser set of preselection cuts is used since the 
lepton energy spectrum is broader due to the increased boost
and the set of variables used in the likelihood selections is modified.
In the \WWqqtn\ sample there is a significant background from
hadronic decays of single W events (\Wenu) and 
an additional likelihood selection is used to reduce this background. 
This is only applied to
\WWqqtn\ events where the tau is identified as decaying in the single prong hadronic 
channel.  
Finally, in order to reduce \Zz \Zz\ contribution, events passing the 
\WWqqen\ likelihood selection are rejected if there is evidence of
a second energetic electron. A similar procedure is applied to the
\WWqqmn\ selection. 
The details of the selection at 183 GeV are given in~\cite{ww183}. 
There is a background of $3.5\%$ from \Zqq\ events,  
$1.0\%$ from \WWqqqq\ events, $1.3\%$ from single W events 
and $0.8\%$ from \ZZqqll\ events. \par

\subsubsection{{\bf Selection of the non-radiative event sample \boldmath{\Zqq }}}
Here, an extension of the selection criteria defined
in~\cite{OPALPR197} is used, which starts by selecting hadronic events defined
as in~\cite{OPALPR035}.  
To reject background from $\epem\rightarrow\tautau$ and
$\gamma\gamma\rightarrow\qq$ and to ensure that the events are
well contained in the OPAL detector one requires that
the event has at least seven charged tracks
with transverse momentum $p_t > 150$~MeV/$c$ and that
the polar angle of the thrust axis lies within the range $|\cos \theta_{T}|<0.9$.
To reject events with large initial-state radiation, one requires
$\sqrt{s}-\rootsprime<10$~GeV, where \rootsprime\ is the effective invariant mass of the hadronic system~\cite{PR183}. 
For the suppression of the \WW\ background one 
requires that the events are 
selected neither for the semileptonic nor for the fully hadronic \WW\ samples described above. 
The cut in the relative likelihood for vetoing \WWqqqq\ events is looser
than in the \WWqqqq\ event selection.
After selection, there is a residual background of $3.8\%$ from W-pair events 
and a contribution of  $0.3\%$ from \Zz \Zz\ events. \par

\subsubsection{Pion selection and Event samples}
\label{pion-sel}
Note that the three event selections result 
in completely independent event samples without any overlap. 
After the event selection the following cuts are applied 
to all tracks, for all three event samples.
A track is required to have a transverse momentum $p_{t} > 0.15$~GeV/$c$,
momentum $p < 10$~GeV/$c$
and a corresponding error of $\sigma_{p} < 0.1$~GeV/$c$. 
Only tracks with polar angles $\theta$ satisfying $|\cos\theta | <
0.94$ are considered. 
The probability for a track to be a pion is enhanced by requiring
that the pion probability P$_{\pi}$ from the d$E$/d$x$ measurement is
P$_{\pi} > 0.02$.
Pion-pairs from a $\mathrm{K^{0}_{S}}$ decay are rejected using the $\mathrm{K^{0}_{S}}$ finder
described in \cite{opalk0k02}. This algorithm rejects 31{\%}
of the unlike-charge pion pairs coming from a $\mathrm{K^{0}_{S}}$ decay.
Since less than $11\%$ of the rejected pairs do not originate from
a $\mathrm{K^{0}_{S}}$, this cut does not introduce a significant bias 
in the $Q$-distribution.
Finally, events with fewer than five charged selected tracks are rejected.
The number of events retained, as well as the number of background events 
evaluated from Monte Carlo simulation is given in
table~\ref{tab-events}, for all three event samples.  \par

\begin{table}[h]
\begin{center}
\begin{tabular}{|c|c|c|c|c|}
\hline
event sample & \multicolumn{2}{c|}{number of selected events} & \multicolumn{2}{c|}{expected background events}\\
 \cline{2-5}
             & 172 GeV &  183 GeV & 172 GeV & 183 GeV \\ \hline
 \WWqqqq & 55 & 327 & $9.5\pm0.5$ & $43.6\pm2.4$  \\
 \WWqqln & 45 & 326 & $2.1\pm0.5$ & $23.1\pm2.4$  \\
 \Zgamma & 214 & 1009 & $8.1\pm1.7$ & $43.2\pm4.9$ \\
\hline
\end{tabular}
\end{center}
\caption{Number of retained events and number of background events predicted for the
  three event samples, separately for 172 GeV and 183 GeV.}
\label{tab-events}
\end{table}

\subsection{Monte Carlo models}
\label{sec-montecarlo}

A number of Monte Carlo models are used to model \Zqq, \WWqqln\ or \WWqqqq\ events.
For the \WWqqqq\ event sample 
the simulated events are also used to determine the fraction of 
track pairs coming from the same or different W bosons.
The Monte Carlo samples are generated at \epm\ center-of-mass energies of 172 and 183 GeV in 
proportion to the corresponding integrated luminosities. 
The production of W-pairs is simulated using \Koralw~\cite{bib-koralw}.  
Non-radiative decays \Zqq\ as well as the \Zz \Zz\ and \singW\ events are 
simulated with \Pythia ~\cite{bib-pythia}. \Koralw\ uses the same string model as \Pythia\ for hadronisation.
For systematic error studies the  event generator \Herwig ~\cite{bib-herwig}, 
which employs a cluster hadronisation model, is also used.
All these Monte Carlo samples discussed above are generated without \bec . 
In addition W-pair events are also simulated with \bec\ included ~\cite{bib-lonnblad}, 
using \Pythia\ \footnote{The model parameters 
controlling \bec\, in \Pythia\ are taken to be 
MSTJ(51)=2, 
MSTJ(54)=--1, 
MSTJ(57)=1, 
PARJ(92)=1.0, 
PARJ(93)=0.4, 
MSTJ(52)=9, 
PARJ(94)=0.275, and 
PARJ(95)=0.0, as suggested by the authors of \cite{bib-lonnblad}.}. 
The algorithm introduces \bec\ via a shift of final-state momenta among 
identical bosons.
For these events two samples are generated: In the first, \bec\, are simulated for all pions 
in the event, both from the same and from different W bosons. In the second sample, \bec\, are 
simulated only for pions originating from the same W boson.

\section{Analysis}
\label{sec-analysis}

\begin{figure}
\begin{center}\mbox{\epsfxsize=16cm
\epsffile{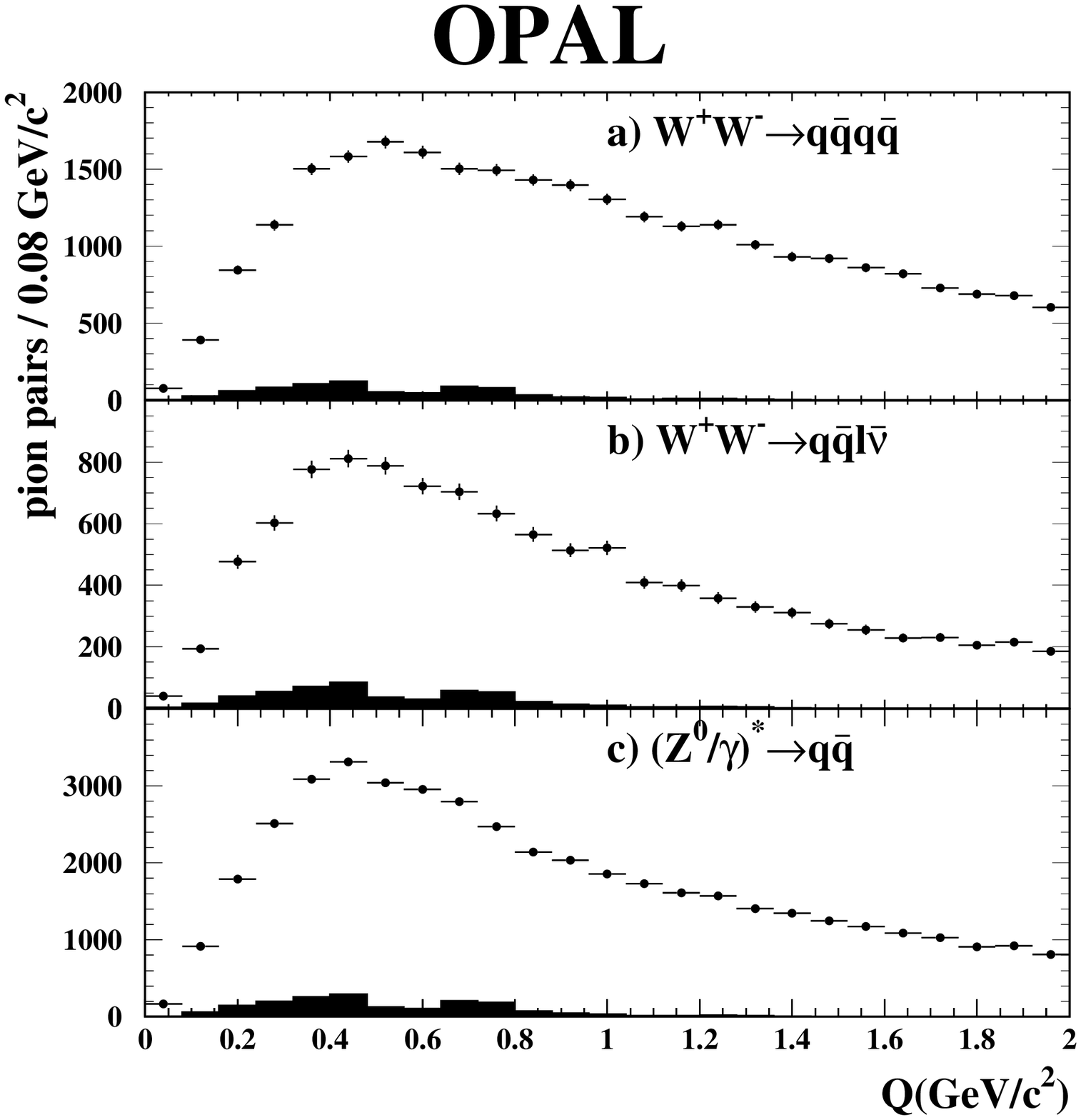}}\end{center}
\caption{The points show the unlike-charge pion pair
distribution, before the correction for resonance production,
for the three different event selections;
a) for the fully hadronic, b) for the semileptonic, and c)
for the non-radiative event selection. 
The filled histogram is the expected contribution from resonances, 
where both tracks of a pion pair come from the same resonance.}
\label{res-cor}
\end{figure}

Using the tracks that pass the selection of section~\ref{pion-sel},
the $Q$-distributions are determined for like-charge pairs as well as 
for unlike-charge pairs. 
The correlation function $C(Q)$ is then obtained as the ratio of these $Q$-distributions.  
Coulomb interactions between charged particles affect like- and unlike-charge 
pairs in opposite ways and modify the correlation function.  We therefore 
apply the following correction to the correlation function,
\begin{equation}
C_{{\rm corr}}(Q) =\chi (Q)\, C_{{\rm uncorr}}(Q),
\label{eq-coulcorr}
\end{equation}
where 
\begin{equation}
\chi (Q)= \frac{\mathrm{e}^{2\pi \eta}\,-\,1}{1\,-\,\mathrm{e}^{-2\pi\eta}},
\end{equation}
and where $\eta = \alpha \, m_{\pi}/Q$ with $\alpha$ the fine-structure constant 
and $m_\pi$  the mass of the charged pion~\cite{bib-coulomb}.  
The Coulomb correction factor $\chi(Q)$ is about 17{\%} in the first $Q$ bin, 
5{\%} in the second bin and 1{\%} in the tenth bin, 
with a bin size of $0.08$ GeV/$c^{2}$ (see Fig.~\ref{res-cor} for the definition of the bins).
The Monte Carlo simulations do not contain Coulomb effects, so the Monte Carlo 
distributions are not corrected by Eq.~\ref{eq-coulcorr}. \par
Structure in the unlike-charge samples due to resonance production
is corrected using Monte Carlo. 
For this, the $Q$-distribution is obtained for unlike-charge pair 
combinations taken exclusively from the decay products of $\mathrm{K^0_S}$ mesons and 
the resonances $\omega$, $\mathrm{\eta}$,
$\mathrm{\eta^{\prime}}$, $\mathrm{\rho^0}$, $\mathrm{f_{0}}$ and $\mathrm{f_{2}}$ 
as produced in the Monte Carlo. 
The production of resonances has only been measured at \epm\ center-of-mass energies around the \Zz\ peak 
and not at energies above the \Zz\ peak.
\Jetset\ \cite{bib-Jetset} describes the production of resonances around the \Zz\ peak 
quite well, although not perfectly in all cases~\cite{had-lafferty}. 
To estimate the contribution for each resonance to the $Q$-distribution 
at LEP 2 energies, the $Q$-distribution for each resonance is multiplied
by the ratio of the measured production rate at LEP \cite{had-lafferty} 
and the corresponding rate in \Jetset. 
The main contributions come from $\mathrm{K^0_S}$, $\omega$, 
$\mathrm{\rho^0}$ and $\mathrm{\eta}$ mesons.
The $Q$-distribution for the
resonances, thus obtained, is then scaled to the number of selected events and
subtracted from the experimental unlike-charge reference $Q$-distribution. 
These corrections are made for each event selection
separately.
They are typically $5-10\%$ for small $Q$, falling 
rapidly for $Q> 0.8$ GeV/$c^{2}$.
The three unlike-charge distributions, 
before the correction, and the expected 
signal from resonance decays are shown in Fig.~\ref{res-cor}.  

\begin{figure}
\begin{center}\mbox{\epsfxsize=16cm
\epsffile{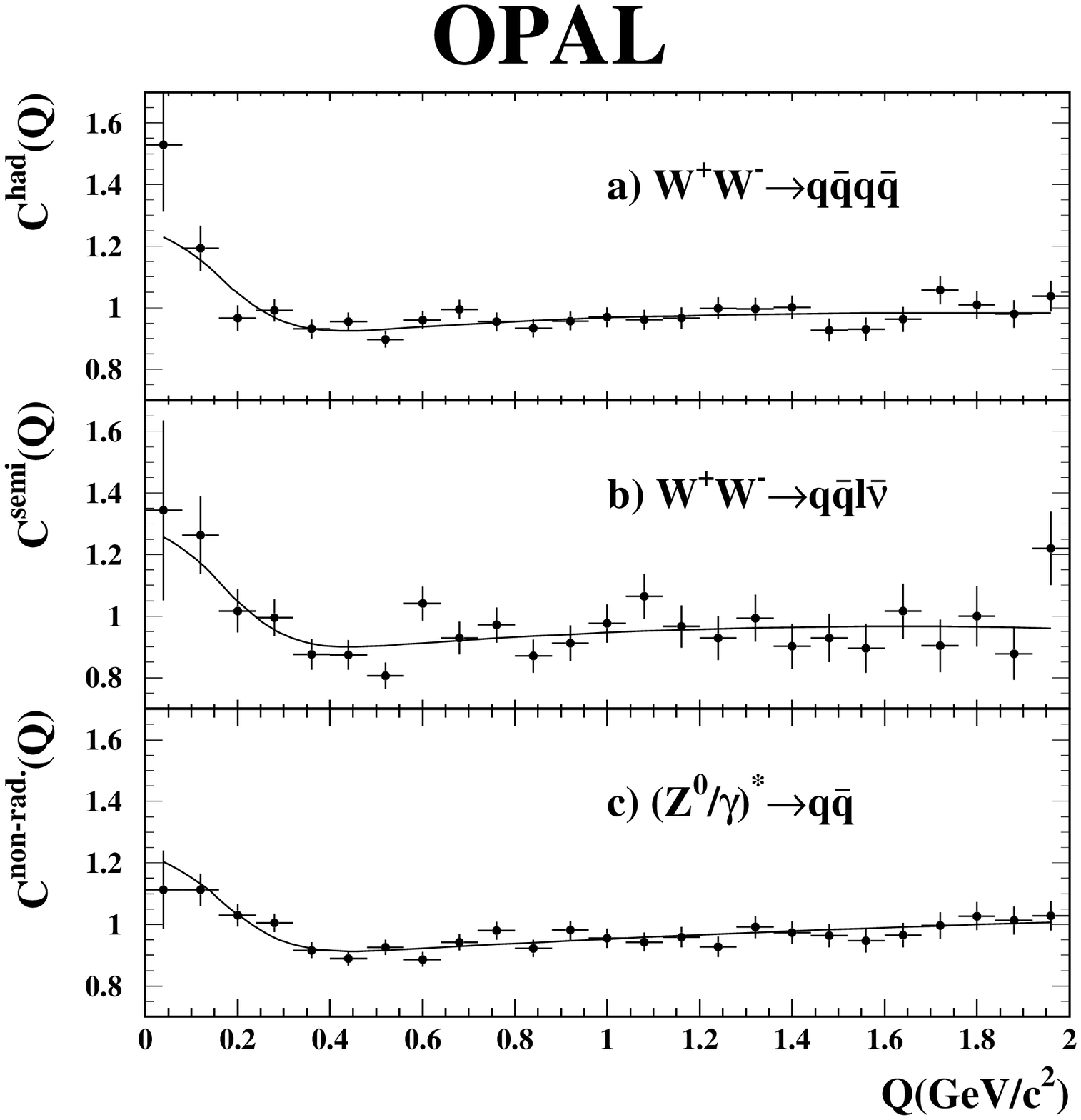}}\end{center}
\caption{
The correlation function for like-charge pairs relative to
unlike-charge  pairs for three event selections;
a) \chad{} for the fully hadronic, b)  \csemi{} for the semileptonic, and c)
\cqcd{} for the non-radiative event selection. 
The Coulomb-corrected data are shown as solid points together with 
statistical errors. 
The curves are the result of the simultaneous fit discussed in
Sect.~\ref{sim-fit}.}
\label{fig-data}
\end{figure}

The resulting experimental correlations $C(Q)$ are shown, for the
three event samples separately, in Fig.~\ref{fig-data}.
The data in all three distributions exhibit a clear enhancement at low
$Q$, consistent with the presence of \bec .

\subsection{\bf Fit to establish \bec\ in W-pair events}
\label{sim-fit-nwa}

The measured distributions cannot be directly  
compared with the parametrisation of Eq.~\ref{eq-usedfun}~ since,  
in general, each distribution has contributions from several
physical processes that may have different \bec . 
To illustrate the situation, consider the hadronic W-pair events.
They have as their main contribution (see Table~\ref{tab-events})
the correlations from pions coming from hadronic W decays. 
They contain, however, also contributions from background events, i.e. \Zqq\ events. 
Thus
\begin{equation}
\chad{} = \frac{N^{\mathrm{WW}}_{\pm\pm} +  N^{\mathrm Z^{*}}_{\pm\pm}}
{N^{\mathrm{WW}}_{+-}  + N^{\mathrm Z^{*}}_{+-}},\label{eq-had-nwa}
\end{equation}
where $N^{\mathrm{WW}}_{\pm\pm}$ and $N^{\mathrm Z^{*}}_{\pm\pm}$ 
are the numbers of like-charge track pairs for the class of pions from
\WWqqqq\ events and for the class of pions from the background sample of \Zqq\ events. The variables 
 $N^{\mathrm{WW}}_{+-}$ and $N^{\mathrm Z^{*}}_{+-}$ are defined analogously for unlike-charge pairs.
Eq.~\ref{eq-had-nwa} can be rewritten as 
\begin{equation}
\chad{}  = \pwwqqqq \,  \cwwqqqq  + \\ 
(1- \pwwqqqq{}) \, \cstarone{} , \label{eq-4q-nwa} 
\end{equation} \\
where \cwwqqqq{}  and \cstarone{} are the \bec\, for the class of pions from
\WWqqqq\ events and for the class of pions from the sample of \Zqq\ events, 
the main background in the hadronic selection.
$\pwwqqqq{}= N^{\mathrm{WW}}_{+-} / (N^{\mathrm{WW}}_{+-} + N^{\mathrm Z^{*}}_{+-})$ 
is the fraction of unlike-charge pion pairs at a given $Q$ which originate from
a W-pair event in the hadronic event sample. 
Here and in the following, the small number of \Zz \Zz 
events are not counted as background but as signal,
since their properties with regard to \bec\ should be quite similar.  \par

The experimentally  determined correlations for the other two
event samples can be written as:
\begin{equation}
\csemi{} = \pwwqqln \, \cwwqqlv +  \\
(1 - \pwwqqln) \, \czstar , \label{eq-semi-nwa}
\end{equation}
for the \WWqqln\ event sample and
\begin{equation}
\cqcd{} = \pznorad  \, \czstar + \\ 
(1 - \pznorad ) \, \cwwqqqq \label{eq-qcd-nwa}
\end{equation}
for the non-radiative \Zgamma event sample. 
The notation in these equations is analogous to that of 
Eq.~\ref{eq-had-nwa}.  \csemi{} and  \czstar{} are the \bec\, for the two 
pion classes from \WWqqln\ and non-radiative \Zqq\ events, respectively. 
The  definition of the relative fractions \pwwqqqq{}, \pwwqqln{} and \pznorad{}
is given in table \ref{tab-defs-nwa}. 
They are taken from a Monte Carlo simulation which does not contain \bec\,  
as discussed in section \ref{sec-montecarlo}. 
These probabilities are global properties of the events and depend 
little on whether \bec\, are assumed or not.  
The small number of single-W events in the semileptonic event sample are treated as signal events. \par

The hadronic W-pair sample contains a sizeable number of \Zgamma
background events. Due to the selection cuts suppressing
\Zgamma events in the hadronic W-pair sample, the remaining \Zgamma events have
different event shapes and multiplicities from those 
in the main non-radiative \Zgamma event sample. Since BEC depend on event shape
and multiplicity~\cite{becmult}, the correlation function for \Zgamma events selected
as hadronic W-pairs, \cstarone{}, is expected to be different from
that for the main non-radiative selection, \czstar{}. To take these
differences into account, the parameters $\lambda$ and $R$ in the
correlation functions \cstarone{} and \czstar{} are not taken to
be equal but those in \cstarone{} are adjusted according to the
different event topology. 
In order to estimate this correction, the \WWqqqq\ selection described in \ref{sec-selection}, 
which contains no direct center-of-mass energy dependent variables, 
is  applied to data  taken at LEP. A simultaneous \bec\ 
fit is applied to both events selected as \WWqqqq\ events and events 
which are not selected as \WWqqqq\ events. The differences obtained
in $\lambda$ and $R$ are used here~\footnote{ For the function \cstarone{} 
the absolute $\lambda$ value is reduced by 0.094 and the 
absolute $R$ value is increased by 0.097 fm relative to the corresponding parameters of \czstar{}, with
\lamzstar{} kept as a free parameter in the main \bec\ fit.} to take 
differences in the correlation function \cstarone{} and \czstar{}
into account.
Due to high purity of the semileptonic and non-radiative \Zgamma\ selections, 
no adjustment is applied to the correlation functions of events selected as background 
in \csemi{} and \cqcd{}. 
For \WWqqqq\ events selected as fully hadronic events and \WWqqqq\ events selected 
as non-radiative \Zgamma events the same \bec\ are assumed.
The effect of this assumption will be described with the systematic errors. \par

\begin{table}
\begin{center}
\begin{tabular}{|c|c|}
\hline
&\\
Probability definition & Prob. that $+-$ track pair   \\
&\\
\hline 
\hline 
&  \\
$\pwwqqqq{} = \frac{N^{\mathrm{WW}}_{+-}(Q)} { N^{\mathrm{WW}}_{+-}(Q) 
  + N^{\mathrm Z^{*}}_{+-}(Q)}$  &  originates from \WWqqqq\ process,\\
& in the hadronic event selection. \\
 &  \\ 
\hline
& \\
\pzstarhad{} = 1 -- \pwwqqqq{}  & originates from \Zqq\  process, \\
& in the hadronic event selection. \\
 &  \\ 
\hline
&  \\
 $\psamesemi{} = \frac {N^{\mathrm{W}}_{+-}(Q)}{ N^{\mathrm{W}}_{+-}(Q)+
N^{\mathrm Z^{*}}_{+-}(Q)}$   &originates from \WWqqln\ process, \\
& in the semileptonic event selection. \\
 &  \\ 
\hline
& \\
$\pznorad{} = \frac{N^{\mathrm Z^{*}}_{+-}(Q)}{N^{\mathrm{WW}}_{+-}(Q) 
  + N^{\mathrm Z^{*}}_{+-}(Q)}$    &originates from \Zqq\ process, \\
&in the non-radiative event selection.  \\
& \\
\hline
& \\
\pwwnorad{} = 1 -- \pznorad{} & originates from \WWqqqq\ process, \\
 &in the non-radiative event selection.  \\
& \\
\hline
& \\
 $\psamehad{} = \frac{N^{\mathrm{ same\,W}}_{+-}(Q)}{ N^{\mathrm{ same\,W}}_{+-}(Q) +N^{\mathrm{ diff\,W}}_{+-}(Q)
  + N^{\mathrm Z^{*}}_{+-}(Q)}$   &originates from the same W, \\
&in the hadronic event selection. \\
 &  \\
\hline
&  \\
$\psameqcd{} = \frac{N^{\mathrm{ same\,W}}_{+-}(Q)}{N^{\mathrm{ same\,W}}_{+-}(Q) +N^{\mathrm{ diff\,W}}_{+-}(Q)
 + N^{\mathrm Z^{*}}_{+-}(Q)}$ & originates from the same W, \\
&in the non-radiative event selection. \\
  &  \\ 

\hline
\end{tabular}
\end{center}
\caption{Definition and meaning of the various probabilities
concerning unlike-charge track pairs, used in 
Eqs. \ref{eq-4q-nwa} - \ref{eq-qcd-nwa} and 
\ref{eq-4q} - \ref{eq-qcd} and illustrated in Figs.~\ref{pur1} and \ref{fig-psame}.}
\label{tab-defs-nwa}
\end{table}

\begin{figure}[ht]
\begin{center}\mbox{\epsfxsize=16cm
\epsffile{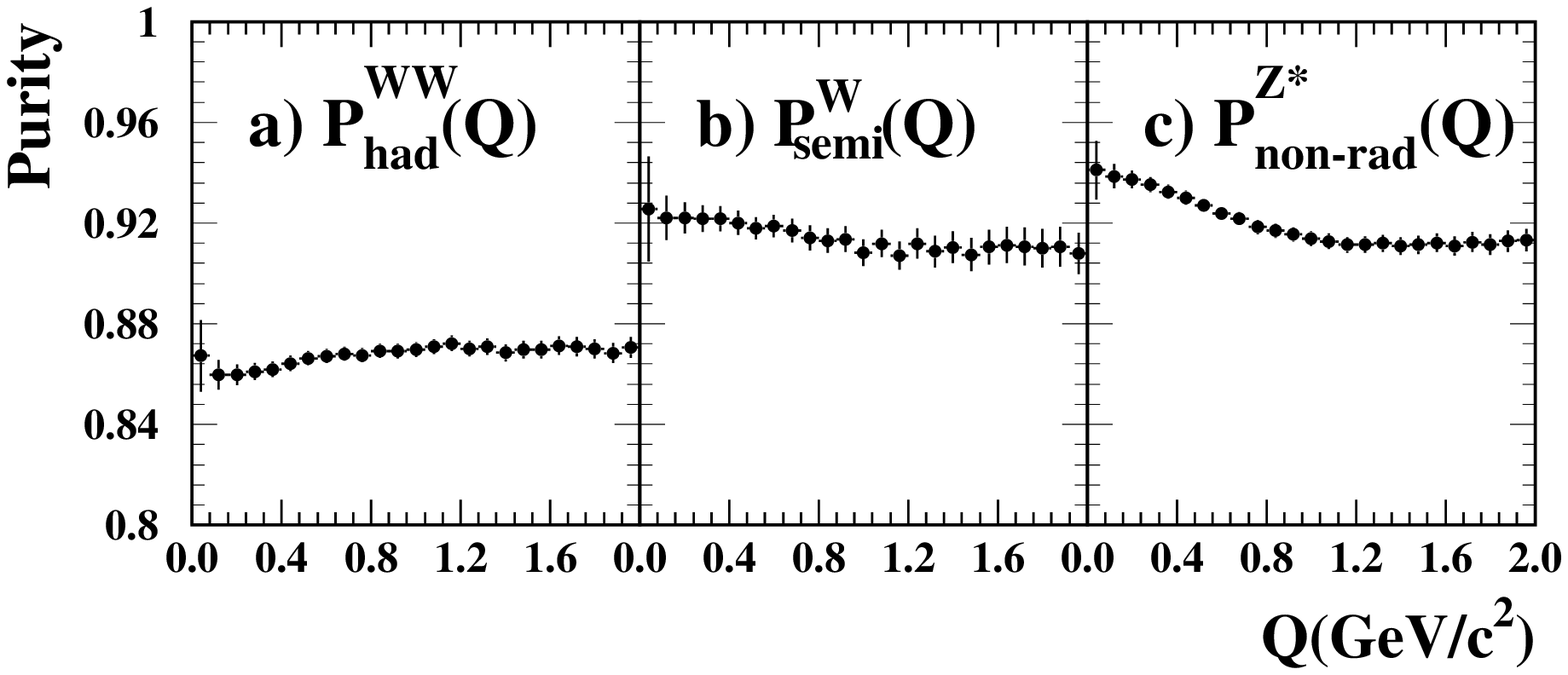}}\end{center}
\caption{The purities a) \pwwqqqq{}, b) \psamesemi{} and c) \pwwnorad{} as obtained
from Monte Carlo simulations.}
\label{pur1}
\end{figure}

The unknown correlation functions \cwwqqqq{}, \cwwqqlv{} and \czstar{} are parametrised 
using  Eq.~\ref{eq-usedfun}.
The parameters are determined in a simultaneous fit to the
three  experimental distributions shown in Fig.~\ref{sim-fit}. 
A common source radius $R$ is used for all event classes, while
the parameter $\lambda$ is allowed to be  different.
There is justified by the typical separation between  the W$^{+}$ and W$^{-}$ decay vertices is smaller than
0.1 fm at LEP 2 energies, much smaller  than the typical hadronic 
source radius of $R \approx 1$ fm~\cite{lund}, justifying equal source radii 
for the \WWqqqq\ and the \WWqqln\ event classes.
The source radius for \epm ~annihilations into hadrons has been measured up to 90 GeV
and no evidence has been found for an energy dependence~\cite{marcellini}.
For this reason $R$ is assumed to be the same at higher energies and the same 
source radius is also used  for the \Zqq\ event class.
Separate fits to the distributions show also consistent radii 
for the different event selections.
The pion probability \fpi\  is taken from Monte Carlo. 
At small values of $Q$ it is about constant $\sim 0.84$ and  varies only weakly with $Q$ for all channels.
The long-range parameters are expected to be different for
the \WWqqqq, \WWqqln\ and \Zqq\ event class, due
to kinematic and topological differences. 
The results for the thirteen free parameters in the fit are given in Table~\ref{tab-results-nwa}.
The fit is made in the full range of $0.0 < Q < 2.0$ GeV/$c^{2}$. 
In the distributions of Fig.~\ref{fig-data} the same particles contribute many
times, in different bins of $Q$,  which introduces bin-to-bin
correlations. These are taken into account in the fit.
All three experimental distributions are well described by the fit, 
 the $\chi^{2}$/d.o.f. is 76.1/62. 

\begin{table}[ht]
\begin{center}
\begin{tabular}{||c||c c c||} \hline
Parameter &   \WWqqqq & \WWqqln & $(\mathrm{Z^{0}}/\gamma)^{*}$  \\ \hline
R (fm) &  & $0.91\pm0.11\pm0.10$  &   \\
$\lambda$  & $0.43\pm0.15\pm0.09$ & $0.75\pm0.26\pm0.18$ & $0.49\pm0.11\pm0.08$\\
$N$ & $0.86\pm0.04\pm0.04$ & $0.79\pm0.08\pm0.08$ & $0.86\pm0.05\pm0.04$ \\
$\delta$ & $0.12\pm0.10\pm0.10$ & $0.29\pm0.23\pm0.24$ & $0.13\pm0.11\pm0.08$  \\
$\epsilon$ & $-0.04\pm0.05\pm0.06$ & $-0.09\pm0.10\pm0.11$ & $-0.02\pm0.05\pm0.04$ \\
\hline
\end{tabular}
\end{center}
\caption{Result of the simultaneous fit.
The first error corresponds to the statistical uncertainty the second to systematics.}
\label{tab-results-nwa}
\end{table}

\subsection{\bf Fit to establish \bec\ in same and different W bosons.}

In this section \bec\ are investigated separately for
pions originating from the same W boson and for 
pions from different W bosons.
The correlations for the fully hadronic event sample 
(Eq. \ref{eq-had-nwa}) are written as 
\begin{equation}
\chad{} = \frac{N^{\mathrm{ same\,W}}_{\pm\pm} + N^{\mathrm{ diff\,W}}_{\pm\pm} +  
 N^{\mathrm Z^{*}}_{\pm\pm}}{N^{\mathrm{ same\,W}}_{+-} +N^{\mathrm{ diff\,W}}_{+-}
+ N^{\mathrm Z^{*}}_{+-}}, \label{eq-hadorig}
\end{equation}
where $N^{\mathrm{ same\,W}}_{\pm\pm}$, $N^{\mathrm{ diff\,W}}_{\pm\pm}$ and $N^{\mathrm Z^{*}}_{\pm\pm}$ 
are the number of like-charge track pairs for the class of pions from
the same W boson, different W bosons and from \Zqq\ events. The variables 
$N^{\mathrm{ same\,W}}_{+-}$, $N^{\mathrm{ diff\,W}}_{+-}$ and $N^{\mathrm Z^{*}}_{+-}$ are defined, 
in a similar way, for unlike-charge pairs.  
Eq.~\ref{eq-hadorig} can be rewritten as 
\begin{eqnarray}
\chad{} = \psamehad{} \, \csame{}  + \pzstarhad{} \, \cstarone{} \nonumber \\ 
+ (1 - \psamehad{}  - \pzstarhad{} ) \, \cdiff{} , 
\label{eq-4q}
\end{eqnarray}
where 
\csame{},\cdiff{}  and \czstar{} are the \bec\, for the class of pions from
the same W boson, different W bosons and from \Zqq\ events. 
The variables $\psamehad{}$ and $\pzstarhad{}$ are defined in Table \ref{tab-defs-nwa}.
Likewise, the experimentally  determined correlations for the other two
event samples can be written as: 
\begin{equation}
\csemi{} = \psamesemi{} \, \csame{} + (1 - \psamesemi{}) \, \czstar{} 
\label{eq-semi}
\end{equation}
for the \WWqqln\ event sample 
and
\begin{eqnarray}
\cqcd{} = \psameqcd{} \, \csame{} + \pzstarqcd{} \, \czstar{}  \nonumber \\ 
+ (1 - \psameqcd{} - \pzstarqcd{} ) \, \cdiff{}
\label{eq-qcd}
\end{eqnarray}
for the non-radiative Z$^{*}$ event sample.
The  definition of the variables
\psamehad{} ,  \pzstarhad{}, \psamesemi{}, \psameqcd{} ,
and \pzstarqcd{} is also given in table \ref{tab-defs-nwa}. 

\label{sim-fit}
By simultaneously fitting  Eq's. \ref{eq-4q},  \ref{eq-semi}, and \ref{eq-qcd}
to the experimental distributions in Fig.~\ref{fig-data}, 
the \bec\ for the three pion classes
\csame{}, \cdiff{}  and \czstar{} are determined.
Again, the probabilities \psameqcd{}, \pzstarqcd{}, \psamehad{},
\pzstarhad{} and \psamesemi{} are taken
from Monte Carlo simulations not containing \bec, as discussed in section \ref{sec-montecarlo}. 

\begin{figure}
\begin{center}\mbox{\epsfxsize=16cm
\epsffile{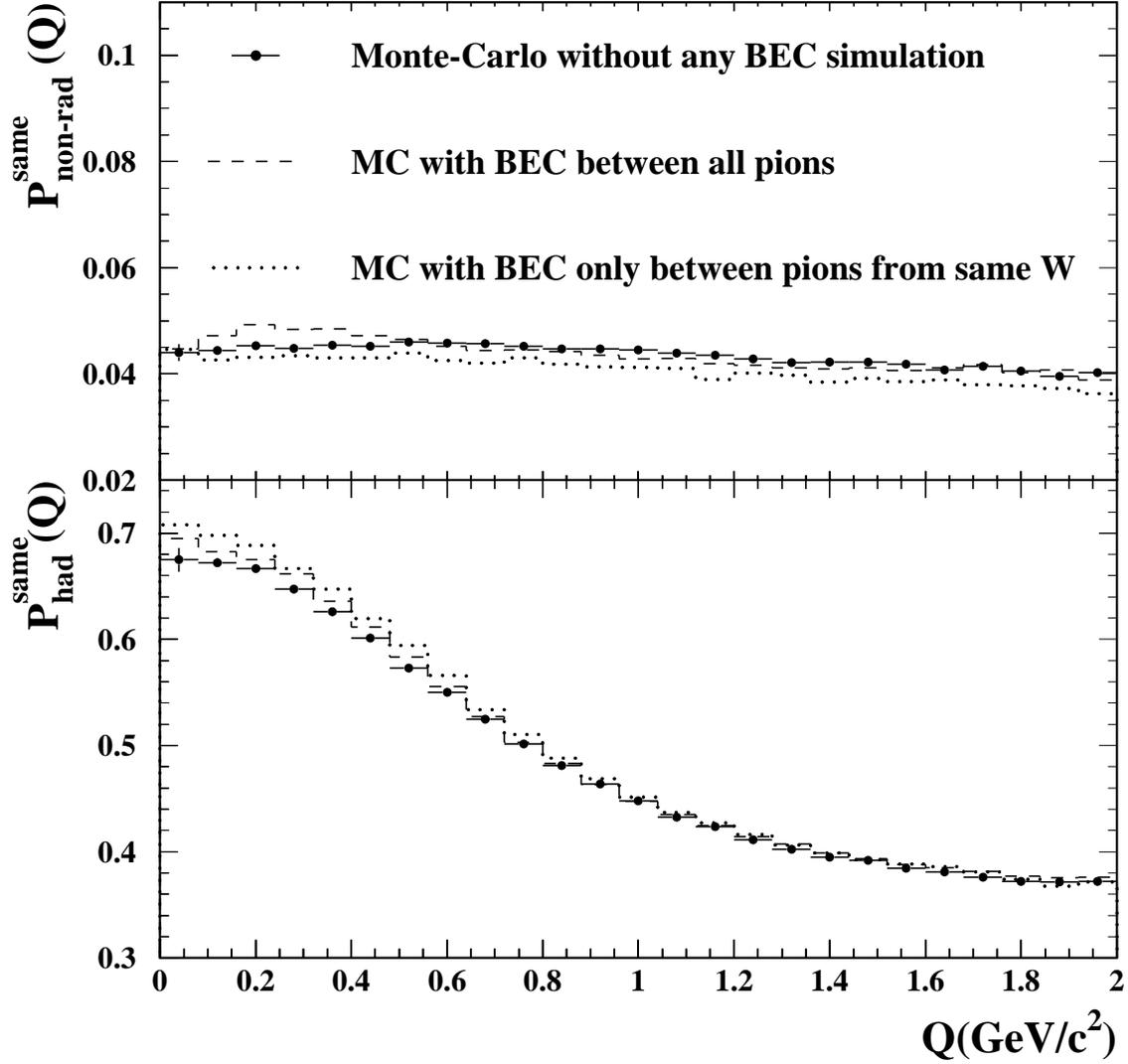}}\end{center}
\caption{
The probability that both tracks of an unlike-charge track pair in
a fully hadronic event originate from the
same W boson, \psameqcd{} (upper plot) and \psamehad{} (lower plot),
 as obtained from Monte Carlo simulations. 
The histogram is the result for the case that no \bec\, are assumed. 
The dashed and dotted histograms are the results for the case that
\bec\, are simulated for all pions or only for pions originating from
the same W boson, respectively. 
}
\label{fig-psame}
\end{figure}

\begin{figure}
\begin{center}\mbox{\epsfxsize=16cm
\epsffile{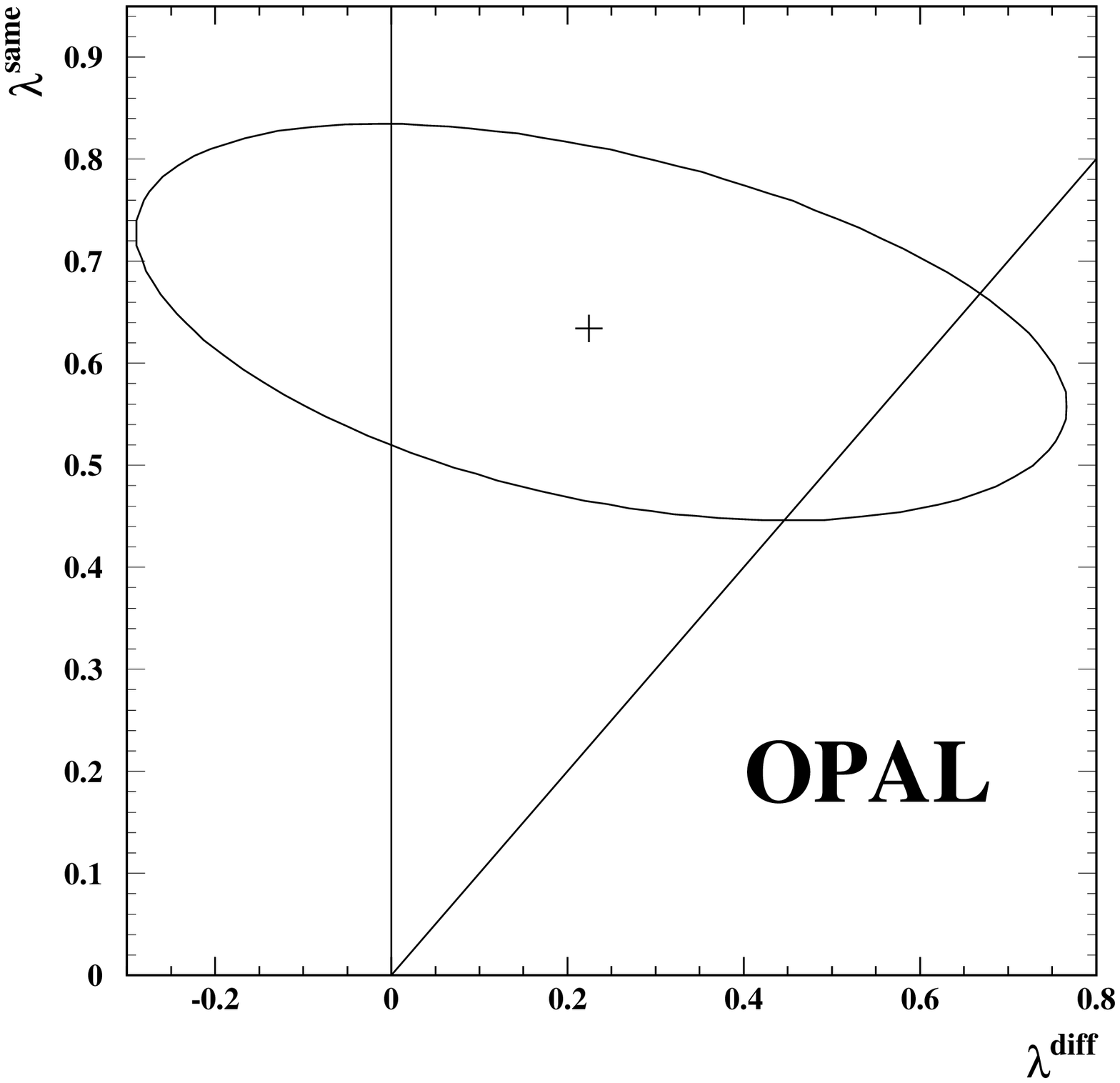}}\end{center}
\caption{Correlation between \lamdiff{} and \lamsame{}. The contour
shows the $67\%$ confidence level. 
The best value obtained in the fit is given by the cross.
The lines for 
 \lamdiff{} = 0
and  \lamdiff{} = \lamsame{} are also indicated.}
\label{contour}
\end{figure}

The functions \psameqcd{} and \psamehad{} are shown in Fig.~\ref{fig-psame}.
Their properties contain only information from unlike-charge pion pairs 
and are therefore independent of \bec.  
The effect of possible variations of the function \psamehad{}, if \bec\, are assumed in the Monte Carlo, 
is discussed in section \ref{sec-systematics}. 
The unknown correlation functions \csame{}, 
\cdiff{}  and \czstar{} are parametrised using  Eq.~\ref{eq-usedfun} 
with different $\lambda$ for the three event classes.  
As before a common source radius $R$ is used for all event classes.
For the correlation function \cstarone{} the parameters $\lambda$ and $R$ are adjusted 
like in section \ref{sim-fit-nwa}. 
Based on Monte Carlo studies the long range parameters \deldiff{} and \epsdiff{}
for the correlation function \cdiff{} are taken to be zero.
This is equivalent to the assumption that colour reconnection effects do not influence the $Q$ distributions. 
The free fit parameters are then determined in a simultaneous fit to the
three  experimental distributions shown in Fig.~\ref{fig-data}. 
The results for the eleven free parameters in the fit are given in Table~\ref{tab-results}.
The fit is made in the full range of $0.0 < Q < 2.0$ GeV/$c^{2}$. 
The fit result is given in Fig.~\ref{fig-data}. 
All three experimental distributions are well described by the fit 
($\chi^{2}$/d.o.f. is 76.4/64). 
The correlation between the parameters \lamdiff{} and \lamsame{}, 
with a coefficient of  $-0.52$,
is shown in Fig.~\ref{contour}.

\begin{table}[ht]
\begin{center}
\begin{tabular}{||c||c c c||} \hline
Parameter & same W & diff W & $(\Zz/\gamma)^{*}$ \\ \hline
$R$ (fm) &  &$0.92\pm0.09\pm0.09$  & \\
$\lambda$ & $0.63\pm0.19\pm0.14$ & $0.22\pm0.53\pm0.14$ & $0.47\pm0.11\pm0.08$ \\
$N$ & $0.83\pm0.05\pm0.07$ & $1.00\pm0.01\pm0.00$ & $0.87\pm0.04\pm0.04$ \\
$\delta$ & $0.21\pm0.15\pm0.19$ & zero assumed & $0.11\pm0.11\pm0.07$ \\
$\epsilon$ & $-0.07\pm0.07\pm0.08$ & zero assumed & $-0.01\pm0.05\pm0.02$ \\
\hline
\end{tabular}
\end{center}
\caption{Result of the simultaneous fit distinguishing pions from the same and different W boson. 
The first error corresponds to the statistical uncertainty, the second one to systematics.} 
\label{tab-results}
\end{table}

\subsection{Systematic Errors}
\label{sec-systematics}
The following variations in the analysis are considered to obtain the systematic error,
which affect the fit result in both fit methods.
The systematic errors are listed 
in Table~\ref{tab-systematics-nwa} and~\ref{tab-systematics},
together with their quadratic sums to give the final systematic error.
\begin{enumerate}

\item {\em Variation of the resonance production.}
In the main analysis, the distortion of the unlike-charge pairs due to resonances was 
taken into account by subtracting the resonance $Q$ distribution 
from unlike-charge pair $Q$ distribution.
This method is based exclusively on \Zz\ data, 
since no measurements of resonance production at LEP 2 are available. 
Thus for systematics  the correction factors
are varied within two standard deviations
of the experimental resonance production cross sections.
The maximum differences in the fit for each resonance were added in quadrature.
Several variations were made for this systematic check, 
therefore no  $\chi^{2}$ is given.
All fits are of good quality.

\item {\em Double-hit resolution.} 
Unlike-charge pairs are bent in a magnetic field in opposite 
directions, whereas like-charge pairs are bent in the same direction. Therefore 
like-charge pairs at very low $Q$ are less well reconstructed. 
Monte Carlo studies indicate the presence of such effects for pairs with a $Q$ less than 0.05 GeV/$c^{2}$. 
For systematics, the fit is repeated in the range between  $0.05 < Q < 2.0$ GeV/$c^{2}$.

\item 
{\em Use of the} \Herwig\ {\em Monte Carlo.}
To determine the purities and the correction for resonances in the unlike-charge 
sample the \Herwig\ Monte Carlo was used.

\item
The {\em probability functions} are obtained from Monte Carlo simulations 
where \bec\, are simulated~\cite{bib-lonnblad} for all pions, both from the same and from
different W bosons.

\item
The {\em probability functions} are obtained from Monte Carlo simulations 
where \bec\, are simulated~\cite{bib-lonnblad} only for pions from the same W.

\item 
{\em  Long-range correlations.}
The fit is repeated with $\epsilon$ = 0.

\item
{\em Different topology of the \Zqq\ background in fully hadronic selected events.}
The difference of the \Zqq\ events in the hadronic and non-radiative Z$^{*}$ samples is taken 
into account in the main  analysis (see section~\ref{sim-fit}). 
The events selected at LEP energies as \WWqqqq\ events and as
non-radiative events, which are used for this correction are statistically limited.
Therefore the parameters
governing the correction factor for the correlation function \cstarone{} 
are varied within their statistical error ($\lambda \pm 0.04$ and $R \pm 0.057$ fm) 
and the largest deviation is taken as the systematic error.

\end{enumerate}

In addition, the effect on uncertainties from arising the knowledge of cross-section is examined. The cross-sections
for W-pair production processes as well as the cross-section for  
non-radiative \Zgamma processes are varied within their experimental uncertainties. 
The impact on the final result is negligible.
Furthermore, differences between \WWqqqq\ events selected as hadronic events and selected as 
non-radiative events are also considered. These variations introduce only small 
changes in the results.

\begin{table}[ht]
\begin{center}
\begin{tabular}{||c||c|c|c|c|c||} \hline
 & $R$ (fm)  & $\lambda^{{\rm \WWqqqq}}$ & $\lambda^{\rm \WWqqln}$ & $\lambda^{{\rm \Zqq}}$& $\chi^{2}/$d.o.f. \\ \hline
 Reference & $0.91\pm0.11$   & $0.43\pm0.15$ &  $0.75\pm0.26$ & $0.49\pm0.11$ & $76.1/62$ \\ \hline
 Variation & $\delta R$ (fm)  & $\delta \lambda^{\WWqqqq}$ & $\delta \lambda^{\WWqqln}$ & $\delta \lambda^{\Zqq }$ &  \\
 \hline
1  & $\pm0.07$ & $\pm0.07$ & $\pm0.10$ & $\pm0.07$ & \\
2  & $<0.01$ & $-0.02$ & $-0.05$ & $+0.03$ & $74.1/62$ \\
3  & $<0.01$ & $+0.03$ & $+0.05$ & $<0.01$ & $95.1/62$ \\
4  & $+0.01$ & $-0.02$ & $-0.02$ & $-0.01$ & $75.7/62$ \\
5  & $<0.01$ & $-0.02$ & $-0.03$ & $<0.01$ & $75.9/62$ \\
6  & $+0.07$ & $-0.03$ & $-0.13$ & $-0.02$ & $78.2/65$ \\ 
7  & $<0.01$ & $<0.01$ & $-0.01$ & $<0.01$ & $76.3/62$ \\ \hline
total & $0.10$ & $0.09$ & $0.18$ & $0.08$ &  \\
\hline
\end{tabular}
\end{center}
\caption{The effect of the systematic variations studied (discussed in
  Sect.~\ref{sec-systematics}) on the variables $R$, \lamwwqqqq{}, \lamwwqqln{}
and \lamzstar{}. The last column shows the quality of the corresponding
fit. 
}
\label{tab-systematics-nwa}
\end{table}

\begin{table}[ht]
\begin{center}
\begin{tabular}{||c||c|c|c|c|c||} \hline
  & $R$ (fm) & $\lambda^{{\rm same}}$ & $\lambda^{{\rm diff}}$ & $\lambda^{Z^{*}}$ & $\chi^{2}/$d.o.f. \\ \hline
 Reference & $0.92\pm0.09$ & $0.63\pm0.19$ & $0.22\pm0.53$ & $0.47\pm0.11$ & $76.4/64$ \\ \hline
 Variation & $\delta R$ (fm) & $\delta \lambda^{\mathrm{same}}$ & $\delta \lambda^{\mathrm{diff}}$ & $\delta \lambda^{\mathrm{Z^{*}}}$ &  \\ \hline
1  & $\pm0.07$ & $\pm0.09$ & $\pm0.07$ & $\pm0.07$ & \\
2  & $<0.01$ & $-0.05$ & $<0.01$ & $+0.03$ & $74.4/64$ \\
3  & $+0.01$ & $+0.04$ & $+0.03$ & $<0.01$ & $94.5/64$ \\
4  & $+0.01$ & $-0.02$ & $-0.10$ & $<0.01$ & $76.0/64$ \\
5  & $<0.00$ & $-0.04$ & $-0.03$ & $<0.01$ & $76.2/64$ \\
6  & $+0.05$ & $-0.08$ & $<0.01$ & $-0.01$ & $77.6/66$ \\
7  & $<0.01$ & $<0.01$ & $-0.05$ & $<0.01$ & $76.5/64$ \\ \hline
Total & $0.09$ & $0.14$ & $0.14$ & $0.08$ & \\ 
\hline
\end{tabular}
\end{center}
\caption{The effect of the systematic variations studied (discussed in
  Sect.~\ref{sec-systematics}) on the variables $R$, \lamsame{}, \lamdiff{} 
and \lamzstar{}. The last column shows the quality of the corresponding
fit. 
}
\label{tab-systematics}
\end{table}

\subsection{Q-based separation of BEC contributions}

\label{sec-unfold}
\begin{figure}
\begin{center}\mbox{\epsfxsize=16cm
\epsffile{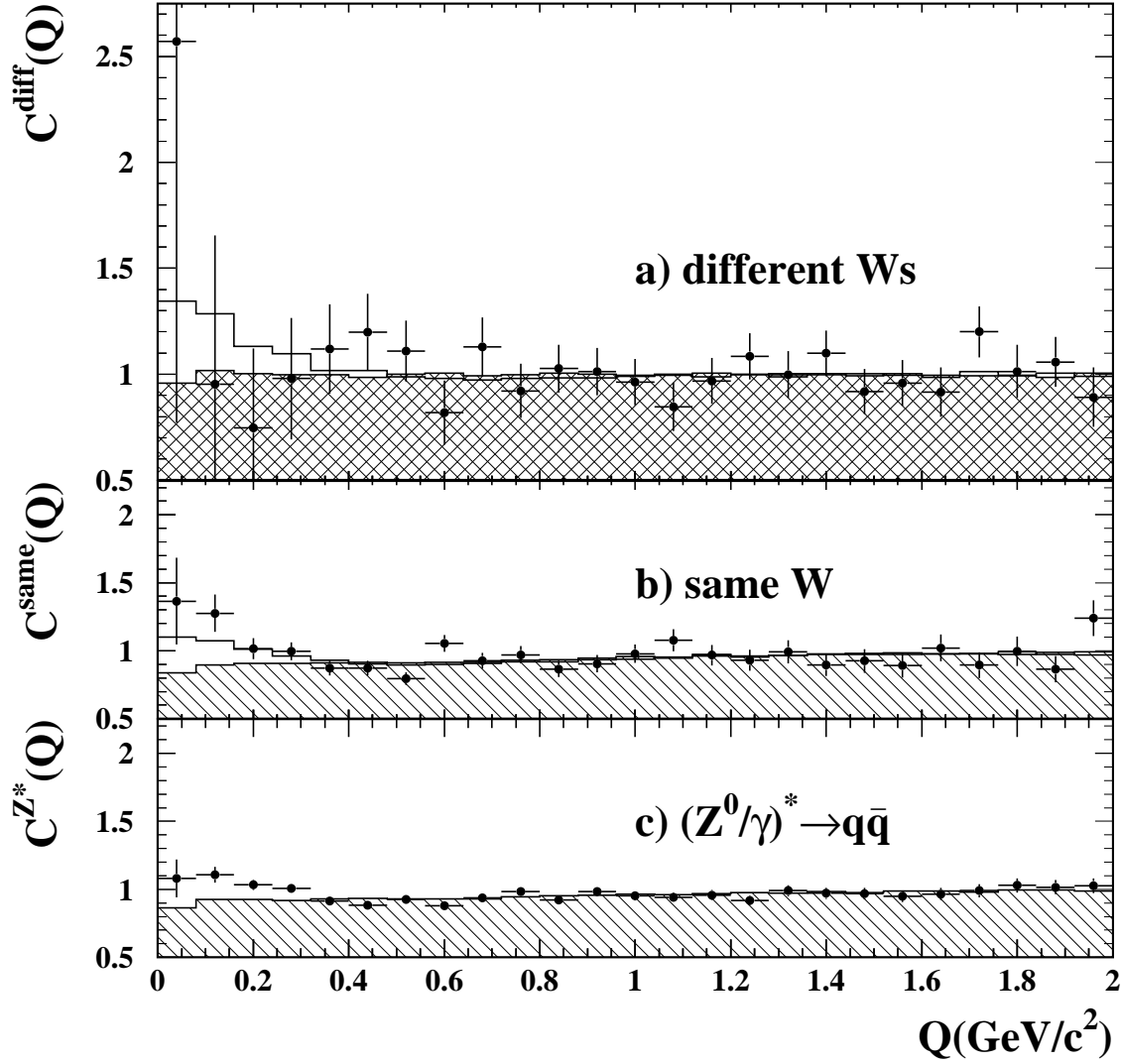}}\end{center}
\caption{Correlation functions for the unfolded classes. The data
 points show the experimental distributions for a pure sample of a)
 pions originating from different W bosons \cdiff{}, b) pions
 originating from the same W boson \csame{} and c) pions from \Zqq{}
 events. The errors are the statistical uncertainties and are correlated 
 between the three classes. The open histogram in a) is the 
 result, for pions from different W bosons,  
 of a simulation including \bec\ between pions from 
  different W bosons, the cross-hatched histogram the corresponding result
 for a simulation with \bec\ for  pions from the same W boson only. 
 The open histogram in b) shows the result, for pions from the same W boson, 
 of a simulation including
 \bec\, for pions from the same W boson and the hatched histogram 
the corresponding result for no \bec\ at all. 
The hatched histogram in c) corresponds to a simulation with no 
 \bec\ at all.}
\label{fig-unfold}
\end{figure}

The experimental \bec\, for pure classes of a) tracks from different
W bosons \cdiff{} , b) tracks from the same W boson \csame{}, as well as
c) tracks from \Zqq\ events \czstar{} can be obtained directly from
Eq.~\ref{eq-4q}~-~\ref{eq-qcd} by solving the equations for these
three unknown functions for each bin of $Q$, using the fractions from Table \ref{tab-defs-nwa}. 
The resulting distributions are shown in Fig.~\ref{fig-unfold}. 
A comparison of data and Monte Carlo without \bec\, shows that there is a clear signal at 
small $Q$ 
for pions originating 
from \Zqq\ events
(Fig.~\ref{fig-unfold}~c). 
The data for pions from the same W boson show a larger enhancement than the corresponding simulation (Fig.~\ref{fig-unfold}~b). 
At the current level of precision, it cannot be established 
whether \bec\, between pions from different W bosons exists or not
(Fig.~\ref{fig-unfold}~a),
in agreement with the result of the simultaneous fit of
Sect.~\ref{sim-fit}. 
Note that the errors of the three unfolded distributions
are highly correlated with each other 
{\em by construction}. 
For $Q$ values larger than about 0.4 GeV/$c^{2}$, the distribution for \cdiff{} is
consistent with being constant in both MC and data. 

\subsection{\bf Consistency check}

\label{sim-fit-dv}

In the analysis described above, the resonances were subtracted using Monte Carlo information 
and long-range correlations were taken into account by the empirical factor 
$(1 + \delta Q + \epsilon Q^2)$ in the correlation function of Eq.~\ref{eq-usedfun}. 
As an alternative, we study here the double ratio
\begin{equation}
C^{\prime} (Q)=\frac{N^{DATA}_{\pm\pm}}{N^{DATA}_{+-}} / \frac{N^{MC}_{\pm\pm}}{N^{MC}_{+-}, }
\end{equation}
where the Monte Carlo events are generated without \bec\ .
If the production of resonances\footnote{As for the main analysis, 
the resonance cross--sections in \Jetset\ are adjusted to the measured rates at LEP energies.}
 and long-range correlations are well described by the simulation, these should cancel in the 
double ratio and only \bec\ should remain. 
The agreement between the simulation and data was checked and is good for both the unlike-charge 
and like-charge distributions. 
The latter show significant deviations only in the low $Q$ region, where distortions 
due to \bec\ are expected in the data. 
Thus, for the double ratio, a simple fit ansatz can be used:

\begin{equation}
C^{\prime} (Q) = N  \, (1 + f_{\pi}(Q)\,\lambda\,
{\mathrm{e}} ^ {-Q^2 R^2}).
\end{equation} 

As in section~3.2, the double ratios for the three event selections can be described 
by superpositions of the correlations for the different pion classes. 
Eqs.~\ref{eq-4q} - \ref{eq-qcd} are also valid for the double ratios. 
It can be shown that the relative probabilities $P$ are given by the expressions given in Table~\ref{tab-defs-nwa}, 
except that, in this case, the number of like-charge pairs $N_{\pm\pm}$ have to be used instead of the number 
of unlike-charge ones $N_{+-}$ as was the case in section 3.2. 
The relative probabilities are determined from a Monte Carlo simulation without \bec . 

In a simultaneous fit to the three double ratios $C^{\prime} (Q)$  the \bec\ for the
three pion classes 
$C^{\prime\,\mathrm{same}}(Q)$, $C^{\prime\,\mathrm{diff}}(Q)$ and $C^{\prime\,\mathrm{Z^{*}}}(Q)$ are determined.  
A common source radius for all 
pion classes is assumed and the  parameters $\lambda$ and $R$ in the
correlation function $C^{\prime\,\mathrm{Z^{*}} }(Q)$  are adjusted for differences in multiplicity and topology as in section 3.2.
The seven parameters used in the fit are given in Table~\ref{tab-results-dv}. 
The fit is made in the full range $0.0 < Q < 2.0$ GeV/$c^{2}$.

The fit describes the distributions well,
with a $\chi^{2}$/d.o.f. of 72.8/67.
The results of the fit are given in Table~\ref{tab-results-dv}. 
They are fully compatible with the results of section 3.2. 
The systematic errors are obtained in a similar way as before,  
with the relevant individual contributions given
in Table~\ref{tab-systematics-dv}. 
This method has the advantage that the long-range correlations
do not have to be determined in the fit.
On the other hand, this method relies more on Monte-Carlo input. 

\begin{table}[ht]
\begin{center}
\begin{tabular}{||c||c c c||} \hline
Parameter & same W & diff W & $(\mathrm{Z}^{0}/\gamma)^{*}$ \\ \hline
R (fm) &  &$1.11\pm0.13\pm0.21$  & \\
$\lambda$ & $0.65\pm0.21\pm0.09$ & $0.50\pm0.78\pm0.14$ & $0.42\pm0.09\pm0.05$ \\
N & $0.99\pm0.01\pm0.03$ & $1.00\pm0.01\pm0.00$ & $0.99\pm0.01\pm0.02$ \\
\hline
\end{tabular}
\end{center}
\caption{Result of the simultaneous fit using the double ratio $C^{\prime}(Q)$
The first error corresponds to the statistical uncertainty the second one to systematics.}
\label{tab-results-dv}
\end{table}

\begin{table}[ht]
\begin{center}
\begin{tabular}{||c||c|c|c|c|c||} \hline
  & $R$ (fm) & $\lambda^{{\rm same}}$ & $\lambda^{{\rm diff}}$ & $\lambda^{Z^{*}}$ & $\chi^{2}/$d.o.f. \\ \hline
 Reference & $1.10\pm0.11$ & $0.64\pm0.20$ & $0.50\pm0.72$ & $0.42\pm0.09$ & $72.8/68$ \\ \hline
 Variation & $\delta R$ (fm) & $\delta \lambda^{\mathrm{same}}$ & $\delta \lambda^{\mathrm{diff}}$ & $\delta \lambda^{\mathrm{Z^{*}}}$ &  \\ \hline
1  & $\pm0.11$ &  $\pm0.07$ & $\pm0.03$ & $\pm0.09$ & \\ 
2  & $-0.14$ & $-0.06$ & $<0.09$ & $0.03$ & $89.6/67$ \\
3  & $-0.12$ & $<0.01$ & $+0.06$ & $+0.03$ & $91.5/67$ \\
7   & $<0.01$ & $<0.01$ & $+0.01$ & $<0.01$ & $72.8/67$ \\ \hline
Total & $0.21$ & $0.09$ & $0.14$ & $0.05$ & \\ 
\hline
\end{tabular}
\end{center}
\caption{The effect of the systematic variations studied (discussed in
  Sect.~\ref{sec-systematics}) on the variables $R$, \lamsame{}, \lamdiff{} 
and \lamzstar{} from the double ratio. 
The last column shows the quality of the corresponding fit. }
\label{tab-systematics-dv}
\end{table}

\section{\bf Discussion and Summary}
We have analysed the data obtained by the OPAL detector at \epm\ 
center-of-mass energies of 172 and 183 GeV to study \bec\ 
between pions in three different physical processes:  
fully hadronic  events \WWqqqq{},  semileptonic events \WWqqln{},  
and  non-radiative \Zgamma events.
The analysis assumes equal source size $R$   
for these processes.
\bec\ are observed each of these processes. 
The chaoticity parameter $\lambda$ for the semileptonic 
process \WWqqln\ is larger than for the processes \WWqqqq\ and \Zqq, 
but still consistent within the errors. 
The long-range correlation parameters are consistent within their errors.
Furthermore, \bec\ between pions from the same W boson and different W bosons 
have been studied.
The result for pions from the same W boson is consistent with those
for pions from non-radiative \Zqq\ events.
At the current level of precision it is not established if 
\bec\ between pions from different W bosons exists or not.

\bigskip\bigskip\bigskip

\par
\section{Acknowledgements:}
\par
We particularly wish to thank the SL Division for the efficient operation
of the LEP accelerator at all energies
 and for their continuing close cooperation with
our experimental group.  We thank our colleagues from CEA, DAPNIA/SPP,
CE-Saclay for their efforts over the years on the time-of-flight and trigger
systems which we continue to use.  In addition to the support staff at our own
institutions we are pleased to acknowledge the  \\
Department of Energy, USA, \\
National Science Foundation, USA, \\
Particle Physics and Astronomy Research Council, UK, \\
Natural Sciences and Engineering Research Council, Canada, \\
Israel Science Foundation, administered by the Israel
Academy of Science and Humanities, \\
Minerva Gesellschaft, \\
Benoziyo Center for High Energy Physics,\\
Japanese Ministry of Education, Science and Culture (the
Monbusho) and a grant under the Monbusho International
Science Research Program,\\
Japanese Society for the Promotion of Science (JSPS),\\
German Israeli Bi-national Science Foundation (GIF), \\
Bundesministerium f\"ur Bildung, Wissenschaft,
Forschung und Technologie, Germany, \\
National Research Council of Canada, \\
Research Corporation, USA,\\
Hungarian Foundation for Scientific Research, OTKA T-016660, 
T023793 and OTKA F-023259.\\

\end{document}